\documentclass[final,5p,times,twocolumn,authoryear]{elsarticle}

\usepackage[utf8]{inputenc}       
\usepackage[T1]{fontenc}          
\DeclareUnicodeCharacter{2212}{-} 

\usepackage{newtxtext,newtxmath}
\usepackage[dvipsnames]{xcolor}
\definecolor{myblue}{RGB}{0,51,102}

\usepackage{hyperref}
\hypersetup{
	colorlinks=true,
	linkcolor=myblue,
	citecolor=myblue,
	urlcolor=myblue
}

\usepackage[pagewise]{lineno}
\usepackage{multirow}
\usepackage{threeparttable}
\usepackage{subcaption}
\usepackage{comment}
\usepackage{adjustbox}
\usepackage{graphicx}
\usepackage{amsmath}
\usepackage{footnote}
\makesavenoteenv{table}
\makesavenoteenv{tabular}
\usepackage{tabularx}
\usepackage{soul}
\usepackage{epstopdf}
\usepackage{float}
\usepackage{lipsum}
\usepackage{pdfpages}

\newcommand{\src}{1A 1246-588, 4U 0614+091 }

\begin{document}
{
\makeatletter
\renewcommand*{\corref}[1]{\@gobble}
\providecommand*{\cnotenum}[1]{}
\makeatother

\begin{frontmatter}
\journal{Journal of High Energy Astrophysics}
\title{Comparative spectro-temporal study between 1A 1246$-$588 and 4U 0614$+$091 using {\em AstroSat}}

\author[1]{Suchismito Chattopadhyay}
\author[2]{Soma Mandal\corref{cor1}}
\ead{soma2778.wbes@gmail.com}
\author[1]{Ranjeev Misra}

\address[1]{Inter University Centre for Astronomy and Astrophysics, Ganeshkhind, Pune 411007, India.}
\address[2]{Government Girls' General Degree College, 7, Mayurbhanj Road, Kolkata 700023, India. }

\begin{abstract}
	We present a comparative broadband spectro-temporal study of the ultra-compact X-ray binary (UCXB) candidates 1A~1246$-$588 and 4U~0614$+$091 using {\em AstroSat} observations. Spectral modelling shows that the spectrum of 1A~1246$-$588 is well described by thermal Comptonization and a soft thermal disk component. We identify a candidate narrow quasi-periodic oscillation (QPO) at $993\pm4$ Hz with a single-detector significance of $\sim3.7\sigma$, tentatively consistent with the upper kHz QPO previously reported from this source. An independent cross-instrument check does not confirm this feature, and we therefore treat it as a tentative, unconfirmed signature rather than a secure detection. In contrast, across the 4 observations, 4U~0614$+$091 evolves from a Comptonization-dominated hard state to a thermally dominated soft state, with the electron temperature decreasing from $\sim19$ keV to $\sim2.3$ keV, while the disk and blackbody temperatures increase. This evolution is accompanied by an increase in the disk contribution to the total flux from $\sim18\%$ to $\sim50\%$ and the appearance of an Fe emission line near 6.97 keV in the soft state. The inner disk radius remains approximately constant at $\sim13$--$15\,R_g$, suggesting that the observed evolution is driven primarily by changes in the thermal properties of the accretion flow rather than substantial variations in disk truncation. When the source is in its hardest state, it displays the richest timing behaviour, with variability components spanning from $\sim12$ Hz to nearly $1$ kHz, whereas the softer observations are dominated by low frequency variability, along with {a} persistent component near $\sim140$ Hz and a prominent $\sim100$ Hz feature in O4. These results suggest that progressive cooling of the Comptonizing corona, along with enhanced thermal emission from the disk and boundary layer, {play a key role} in shaping the spectral and timing evolution of 4U~0614$+$091.
\end{abstract}


\end{frontmatter}
}
\section{Introduction}
\label{sec:intro}

In neutron star low mass X–ray binaries (NSLMXBs), matter from a companion star $\leq 1 M_{\odot}$ gets transferred to the compact object, a neutron star (NS), through the inner Lagrange point \citep{klitzing:frank1985accretion}. These systems feature orbital periods ranging from longer than an hour to several hundred days. On the other hand, ultra-compact X-ray binaries (UCXBs); a subclass of LMXBs, have an orbital period shorter ($<$ 80 minute) in comparison to NSLMXBs. In such short-period binaries, the components must be so close together that ordinary, hydrogen-rich stars do not fit in \citep{nelson86}. Both stars in ultra-compact binaries must thus be compact stars: the cores of evolved giants, white dwarfs (WDs), NSs, or black holes (BHs). The observed helium and carbon-oxygen composition of the transferred matter confirms that a WD or a helium-burning star is the donor \citep{Nelemans2004, Nelemans2006}.

Since companion stars for UCXBs are chemically distinct from a main-sequence star in NSLMXBs, the abundances in the accretion disk also differ quite significantly from those of non-ultra-compact LMXBs. UCXBs provide unique laboratories to study accretion processes in hydrogen-deficient environments. Apart from being useful avenues for studying accretion physics and probing the nature of compact objects, these are expected to play a crucial role in the new era of gravitational wave astronomy, as they fall into the category of well-known persistent sources. Hence, they are predicted to be detectable by future multi-messenger missions such as LISA \citep{Nelemans_2009, Chen-2020, 2023LRR, Chen-2025}, TianQin \citep{Li2025} and Taiji \citep{Ruan2020}. Hence, understanding and classifying these systems well in advance is useful.

Both classes exhibit rich spectral and temporal variability over a wide range of timescales \citep{2003A&A...405..199S,2006A&A...460..229S,2025A&A...695A..44D,2026A&A...710A..43B}, but broadband, simultaneous spectro-temporal characterisation of individual UCXBs remains comparatively limited \citep{2019ApJ...883...39L}, since previous studies of many of these sources have generally relied on narrower-band or non-simultaneous soft/hard X-ray coverage. In this work, we exploit the broad, simultaneous energy coverage of \textit{AstroSat} to jointly constrain the spectral and fast-timing properties of two UCXB candidates, 1A~1246$-$588 and 4U~0614$+$091.

\begin{table*}[h]
\renewcommand{\arraystretch}{1.1}
\centering
\caption{Details of the {\em AstroSat} observations}
\label{tab:observations}
\small
\setlength{\tabcolsep}{16pt}
\begin{tabular}{lllll}
\hline
Source & Observation ID & Observation Date & Exposure Time & Name \\
& & & (LAXPC) & \\
\hline
1A 1246$-$588   & G07\_065T02\_9000001150 & 8--9 Apr, 2017     & 46 ks & O1\\
\hline
4U 0614$+$091   & A02\_175T01\_9000000874 & 10 Dec, 2016       & 13 ks & O2\\
               & A02\_175T01\_9000000976 & 23--24 Jan, 2017   & 10.6 ks  & O3\\
               & G06\_036T01\_9000000988 & 28 Jan, 2017       & 29 ks & O4\\
               & A04\_017T01\_9000001932 & 01--02 Mar, 2018   & 30 ks  & O5\\
\hline
\end{tabular}
\label{bpl}
\end{table*}

1A~1246$-$588 was discovered in 1976 with the \textit{Ariel-V} mission \citep{1976MNRAS.177P..13S}. The first pointed observation was obtained with \textit{EXOSAT} in 1985. A Type-I X-ray burst observed with \textit{BeppoSAX} \citep{Piro:1997IAUC.6538....2P} identified 1A~1246$-$588 as an NSLMXB, and several intermediately long bursts from this source have since been reported \citep{IntZand2008:2008A&A...485..183I}. \citet{bassa2006} identified the optical counterpart and established the UCXB nature of the system through the X-ray-to-optical flux ratio. From a radius-expansion burst, \citet{IntZand2008:2008A&A...485..183I} derive a distance of $\approx 4.3$~kpc, placing 1A~1246$-$588 among the persistently faint LMXBs, with a 2--10~keV luminosity of $\sim 2 \times 10^{35}\,\mathrm{erg\,s^{-1}}$. The \textit{RXTE} all-sky-monitor (ASM) has persistently detected the source in X-rays, and it is notably one of the weakest LMXBs for which a high-amplitude, highly coherent kilohertz QPO at $1258$~Hz has been reported using \textit{RXTE} data \citep{Jonker-2007/10.1111/j.1365-2966.2007.11854.x}. No disc-blackbody component could be robustly identified in earlier broadband spectral fits of this source \citep{IntZand2008:2008A&A...485..183I}, leaving the disc geometry and its connection to the kHz QPO phenomenology in this faint UCXB an open question that a simultaneous soft- and hard-X-ray spectral-timing analysis can help address.

4U~0614$+$091, a confirmed UCXB \citep{2023A&A...677A.186A}, was detected in 1972 \citep{Giacconi1972:1972ApJ...178..281G}, with an orbital period of 51.3~min \citep{Shahbaz:2008PASP..120..848S} and at a distance of 3.2~kpc \citep{Galloway2020:2020ApJS..249...32G}; the orbital period was measured directly from modulations in the optical light curve driven by the uneven heating of the donor star. Type-I X-ray bursts have been detected in this system \citep{Lewin:1976IAUC.2914....3L,kuulkers05:2005ATel..483....1K,kuulkers10:2010A&A...514A..65K}, confirming the compact accretor as an NS \citep{Brandt:1992A&A...262L..15B}, while the presence of oxygen and neon features indicates a WD donor \citep{Juett:2001ApJ...560L..59J, Nelemans2004}. Burst oscillations have established a spin frequency of 414.7~Hz \citep{Stroh:2008ApJ...672L..37S}. Combined radio-to-X-ray spectral data have confirmed the presence of a radio counterpart and characterised the associated jet \citep{Migliari:2010ApJ...710..117M}. Previous spectral studies have reported both O\,{\sc viii} and Fe~K reflection features in this system and found a subsolar Fe abundance \citep{Madej10:2010MNRAS.407L..11M, Madej13:2013MNRAS.429.2986M, ludlam19:2019ApJ...873...99L, ludlam:2024ApJ...975...68M}. Despite this comparatively rich spectral characterisation, these reflection studies have generally relied on models calibrated for solar-abundance, hydrogen-rich discs, leaving open the extent to which oxygen-rich disc composition affects the reliability of standard reflection diagnostics in this source, and whether the fast-timing behaviour tracks the spectral-state changes inferred from these models.

Motivated by these open questions, we use the broad, simultaneous energy coverage of \textit{AstroSat} (spanning the soft and hard X-ray bands via SXT and LAXPC, respectively) to jointly constrain the spectral components and characterise the rapid X-ray variability of 1A~1246$-$588 and 4U~0614$+$091, with the aim of testing whether their broadband spectral and timing properties are self-consistent within a common disc-corona framework, and of examining the disc and reflection diagnostics of each source under simultaneous soft- and hard-band constraints not available to several of the earlier, narrower-band studies discussed above.

The article is organised as follows. The observations and data reduction procedures are described in Section~\ref{sec:reduction}. Section \ref{sec:hid} explains the evolution of the source in their respective HIDs. The timing and spectral results are presented in Sections~\ref{sec:temp} and \ref{sec:spec}, respectively. Finally, the implications of our results are discussed in the summary and conclusions.

\section{Observations \& Data reduction}
\label{sec:reduction}

\textit{AstroSat} has observed the source 1A 1246-588 only once during April 8-9, 2017, whereas 4U 0614+091 has been observed on multiple occasions. The details of the observations are mentioned in Table \ref{bpl}. To analyze the LAXPC data the standard software {\tt LAXPCsoftware22Aug15} \footnote{\url{http://astrosat-ssc.iucaa.in/laxpcData}} has been extensively used. For each observation, we have merged all the orbit files to create a single level 2 file from which the advanced scientific products have been generated using the standard (free from South Atlantic Anomaly (SAA) and earth occultation) good time intervals (GTI). It is to be noted that for the temporal analysis we have made use of all the proportional counter units (PCUs) if the observations belong to 2016 or 2017 while for the case of observations from 2018 we only use the PCU 20 for the further analysis as the PCU 30 switched off due to gain-related issues, and PCU 10 has been reported to have a higher background count due to the functional issue of one of its veto anodes \citep{klitzing:2017ApJS..231...10A}. For the spectral analysis, only PCU 20 data has been used throughout, due to its lower background noise.

{\tt SXTPIPELINE}\footnote{\url{https://www.tifr.res.in/~astrosat\_sxt/sxtpipeline.html}} has been used to process the SXT level1 data to produce the level2 data. Level2 data from different orbits have been merged using {\tt SXTEVTMERGERTOOL} (Julia-based module)\footnote{\url{http://astrosat-ssc.iucaa.in/sxtData}} to produce the merged event file, which is further analysed using {\tt HEAsoft} v6.32 routine {\tt XSELECT}. The count rates for all cases are found to be well below 40 cts s$^{-1}$, and therefore, no pile-up correction has been employed for the analysis. A region of 720 arcsec has been used to produce the SXT spectra and light curves.

\section{HID evolution}
\label{sec:hid}

The hardness-intensity diagram (HID) is constructed by defining the hardness ratio (HR) as the ratio of count rates in the hard energy band (8--20 keV) to those in the soft energy band (3--8 keV), and plotting HR on the \textit{y}-axis. The \textit{x}-axis represents the intensity, calculated as the sum of count rates in both energy bands. In all cases, we have only chosen PCU 20 data to construct the HID for further comparisons.

\subsection{1A 1246-588}
\label{sec:hid_1a1246}

\begin{figure}[ht]
\includegraphics[width=0.45\textwidth]{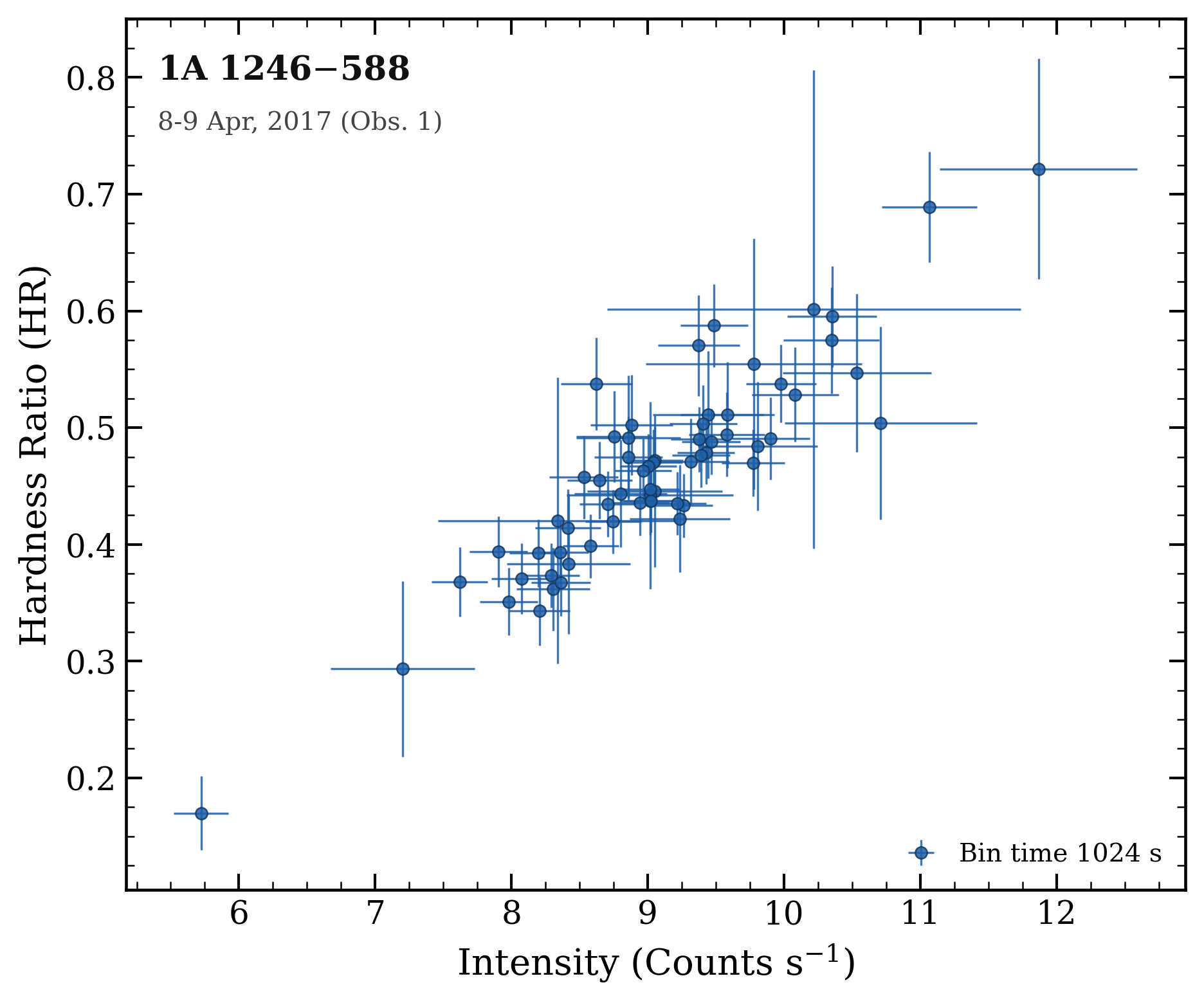}
\caption{Long-term Hardness-Ratio (HR) vs Intensity (I) evolution of 1A 1246-588 using {\em AstroSat} PCU 20. The \textit{y}-axis represents the HR, which is defined as the ratio of the background-subtracted count-rates in the 8-20 keV energy band to the 3-8 keV energy band, while the \textit{x}-axis gives the sum of the count rates in these two different bands.}
\label{Fig0_p2}
\end{figure}

In this case, during the single observation, the source exhibits a relatively broad intensity range, varying from 6 to 12 counts s$^{-1}$, with the HR fluctuating between 0.2 and 0.8. The low intensity likely indicates that the source is intrinsically faint (see Figure \ref{Fig0_p2}). \citet{Poojyam2026} showed that the HID occupies a region of relatively low intensity and high hardness. This comparatively hard location in the HID is primarily a consequence of the adopted hardness definition, which emphasises the energy range where the Comptonized component contributes substantially, even when the source is classified as being in the soft state based on its broadband spectral decomposition.

\subsection{4U 0614+091}
\label{sec:hid_4u0614}

\begin{figure}[ht]
\includegraphics[width=0.5\textwidth]{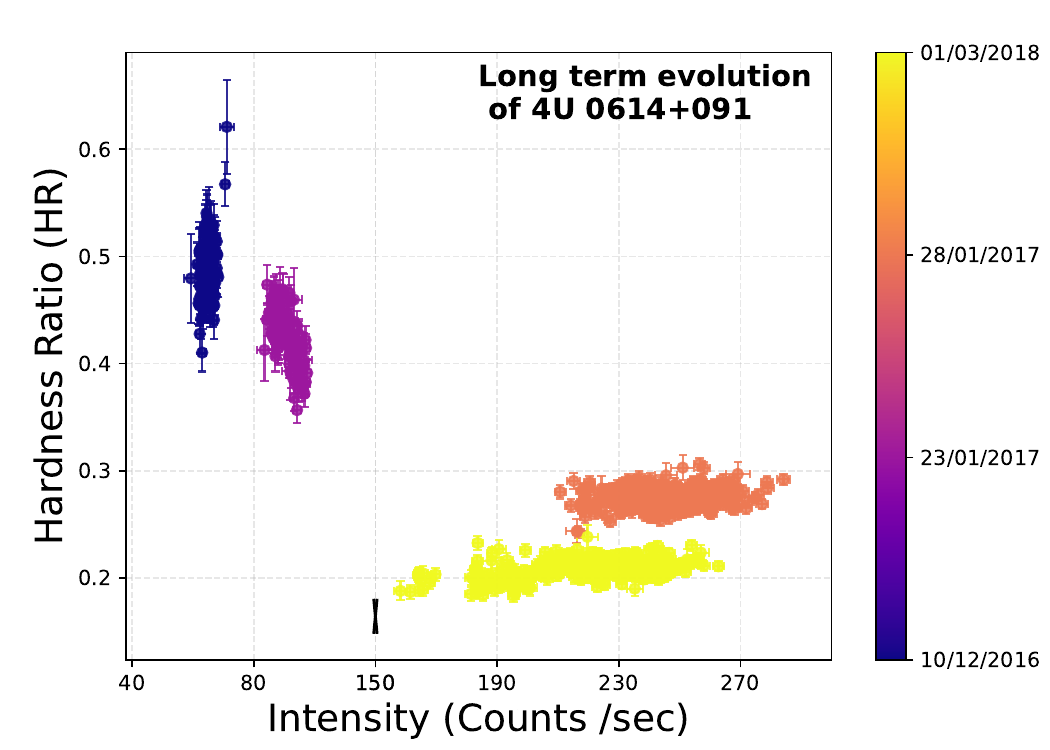}
\caption{Long-term Hardness-Ratio (HR) vs Intensity (I) evolution of 4U 0614 + 091 using {\em AstroSat} PCU 20. The \textit{y}-axis represents the HR, which is defined as the ratio of the background-subtracted count-rates in the 8-20 keV energy band to the 3-8 keV energy band, while the \textit{x}-axis gives the sum of the count rates in these two different bands.}
\label{Fig0}
\end{figure}

As shown in Figure~\ref{Fig0}, two of the selected observations correspond to relatively hard spectral states, while the other two indicate softer states, characterised by an increased count rate from approximately 40--80 counts~s$^{-1}$ to about 150--270 counts~s$^{-1}$. Although the softer states exhibit extended regions that could be divided into multiple zones to explore the system's evolution in more detail, we chose to focus on overall state transitions and therefore did not further subdivide these regions.

\section{Timing Analysis}
\label{sec:temp}

The routine {\em{laxpc\_find\_freqlag}} has been used for the creation of the PDS. In each scenario, the Nyquist frequency is set to 2000 Hz, and the PDS resolution is determined by the segment length used to create the PDS. It is to be noted that all the PDS are dead-time corrected and Poisson noise subtracted. The PDS is then fitted with multiple Lorentzian components and a constant to account for faint residuals. The best-fitting parameters are tabulated in Tables \ref{Table_pds} and \ref{Table_pds2}.

\subsection{1A 1246-588}
\begin{figure}[h]
    \centering
    \includegraphics[width=0.45\textwidth,height=0.3\textheight]{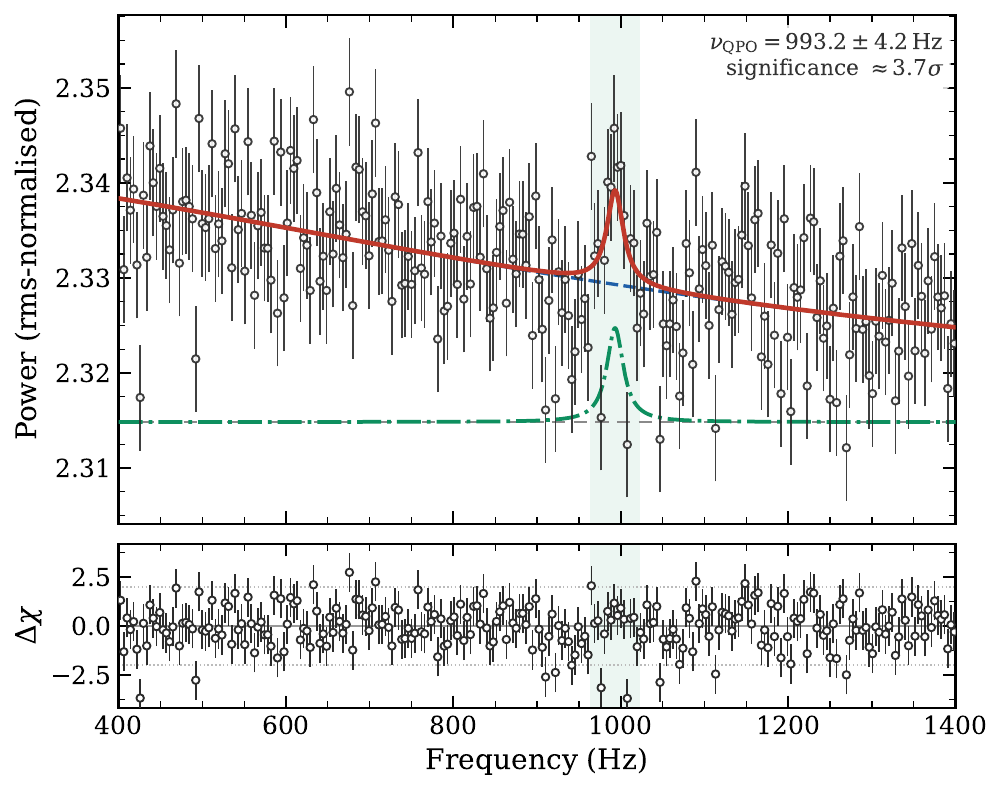}
    \caption{Power density spectrum (PDS) of the source 1A 1246-588 using the LAXPC 20 within 3-20 keV energy range. The bin time used to create the PDS is 0.00025 seconds, and the PDS is fitted with multiple Lorentzians and a Constant component to account for the noise component. Significance of the kHz QPO is $\sim$ 3.7 $\sigma$ with a centroid frequency 993 $\pm$ 4.2 Hz. This feature was not independently confirmed by a cross-instrument check; see Section~\ref{sec:temp}.}
   \label{fig:pds_1A1246}
\end{figure}

In the case of Observation 1, we observed the presence of a marginal kHz QPO around 993 Hz, with a significance of $\sim$ 3.7 $\sigma$ (see Figure~\ref{fig:pds_1A1246}). The significance is defined as the ratio of the normalisation and its negative error. It is further to be noted that the PDS were rms-normalised, for which the normalisation corresponds to the squared fractional rms of the variability feature. For the detected QPO, the fitted Lorentzian normalisation of $N = 0.37 \pm 0.10$\color{black}\ implies an observed fractional rms of
\begin{equation}
\mathrm{rms}_{\mathrm{obs}} = \sqrt{N} = 0.61 \pm 0.08\color{black}\ .
\end{equation}

The LAXPC data are strongly background-dominated, with the source contributing only $\sim 16\%$ of the total count rate. In this background-dominated regime, the Poisson noise level in the rms-normalised PDS is itself elevated, following $P_{\mathrm{noise}} = 2(S+B)/S^2$; for the present rates ($S \approx 5.9$ cts s$^{-1}$, $S+B \approx 40.2$ cts s$^{-1}$) this gives $P_{\mathrm{noise}} \approx 2.3$, consistent with the fitted constant in Table~\ref{Table_pds}.\color{black}\ In principle, the intrinsic rms can be estimated by correcting for background dilution using the method outlined by \citet{vanderklis1989}
\begin{equation}
\mathrm{rms}_{\mathrm{corr}} = \mathrm{rms}_{\mathrm{obs}} \times \frac{S+B}{S},
\end{equation}
which yields $\mathrm{rms}_{\mathrm{corr}} \approx 3.80 \pm 0.52\color{black}$. This uncertainty combines the statistical error on $\mathrm{rms}_{\mathrm{obs}}$ ($\sim$13.5\%) in quadrature with the uncertainty on the background-subtracted source rate obtained from the corresponding spectral extraction, $S = 5.817 \pm 0.122$ cts s$^{-1}$ ($\sim$2.1\%), which self-consistently incorporates the Poisson counting uncertainty of both the source and background spectra. Since the QPO normalisation uncertainty dominates the error budget, the background-rate uncertainty contributes only marginally to the total, and the conclusion is robust to this additional source of uncertainty.\color{black}\ Such a value is unphysically large, and reflects the low source fraction in the LAXPC band. Therefore, background-corrected rms amplitudes are not physically meaningful in this case. This unphysical increase of fractional RMS due to low signal-to-background ratios aligns with the statistical limits documented by \citet{vaughan2003}. We hence quote the observed fractional rms derived directly from the Lorentzian normalisation, noting that background dominance primarily increases the statistical uncertainty without biasing the QPO detection or centroid frequency. We note that 1A~1246$-$588 is itself among the persistently faintest LMXBs known (Section~\ref{sec:intro}), with a 2--10~keV luminosity of only $\sim2\times10^{35}\,\mathrm{erg\,s^{-1}}$; this intrinsic faintness is the underlying reason the source contributes so small a fraction of the total LAXPC count rate, and is therefore a plausible contributing factor, alongside the background-dominated regime discussed above, to any residual bias in the single-detector noise modelling.

	As an additional independent check, motivated by this concern, we calculated the cross-power spectrum (co-spectrum) between the LAXPC10 and LAXPC20 event lists for Observation~1. We used the same good time intervals and 3--20~keV energy range as in the single-detector analysis. The segment length was chosen to match exactly the frequency resolution of the PDS discussed above (0.256~s, corresponding to $\Delta\nu = 3.90625$~Hz). Since the Poisson noise and dead-time effects in the two detector units are statistically independent, their contribution to the expected cross spectrum is zero. Therefore, unlike the single-detector PDS, the co-spectrum does not require a noise-model subtraction and is not affected by the background-related bias discussed above.
	
	We fitted the co-spectrum using the same Lorentzian-plus-constant model, keeping the centroid frequency and FWHM fixed at the values obtained above (993.2~Hz and $<12$~Hz, respectively). The resulting Lorentzian normalisation is consistent with zero, with a significance of $<0.01\sigma$. Expressed in the same normalisation convention as $N$ above, we obtain $N_{\mathrm{cospec}} \approx 0.0013 \pm 0.22$, which is in mild ($\sim1.7\sigma$) tension with the value obtained from the single-detector PDS. The uncertainty in the co-spectrum is substantially larger than that of the single-detector PDS. This is expected because LAXPC10 has a higher background than LAXPC20 \citep{klitzing:2017ApJS..231...10A}. The higher background increases the statistical uncertainty of the cross term, but does not introduce a systematic bias in the measurement. Thus, this cross-instrument test does not provide independent confirmation of the candidate QPO, but neither does it conclusively rule it out. Considering this result together with the already marginal significance of the feature in the single-detector PDS and the possibility of residual background-related bias in the normalisation discussed above, we do not regard this feature as a secure QPO detection. We therefore refer to it as a candidate signature throughout the remainder of this work.

\begin{table}[ht]
    \centering
        \caption{The best fitted value of the PDS in different observations using the multiple Lorentzians along with a constant component. $\dagger$ sign represents the parameters that have been held fixed during the PDS fitting. $\nu$ represents the centroid frequency, FWHM represents the full width at half maximum, and Norm represents the normalisation of each Lorentzian. DOF stands for degrees of freedom. Lorentzian~2 in O1 corresponds to a candidate feature that could not be independently confirmed via a cross-instrument check (Section~\ref{sec:temp}).}
    \begin{tabular}{c c c}
    \hline
         Component & Parameters & O1  \\
    \hline
    {\texttt{Const}} & & 2.31 $\pm$ 0.007 \\
    \hline
    {\texttt{Lorentzian 1}} & $\nu_0$ (Hz) & 0$^{\dagger}$  \\
                            &  FWHM$_0$ (Hz) & 1077 $\pm$ 406  \\
                            & Norm$_0$ & 90.7 $\pm$ 50.5   \\
    \hline
    {\texttt{Lorentzian 2}} & $\nu_1$ (Hz) & 993 $\pm$ 4.2  \\
                            & FWHM$_1$ (Hz) & < 12  \\
                            & Norm$_1$ & 0.37 $\pm$ 0.1    \\
    \hline
    {\texttt{Reduced $\chi^2$}}/DOF & & 1.18 / 251  \\
    \hline
    \end{tabular}
    \label{Table_pds}
\end{table}

\begin{table*}[ht]
\renewcommand{\arraystretch}{1.2}
\centering
\caption{
Best-fitting parameters of the power density spectrum for Observations O2--O5 using multiple Lorentzian components and a constant.
$\nu$ denotes the centroid frequency, FWHM the full width at half maximum, and Norm the Lorentzian normalisation expressed in units of $10^{-2}$.
The $\dagger$ symbol indicates parameters that were held fixed during the fit.
}
\begin{tabular}{c c c c c c}
\hline
Component & Parameters & O2 & O3 & O4 & O5\\
\hline

{\texttt{Const}}
& Norm $\times10^{-4}$
& $3.9 \pm 0.2$ & $1.79 \pm 0.6$ & $0.39 \pm 0.016$ & $0.74 \pm 0.063$\\
\hline

{\texttt{Lorentzian 1}}
& $\nu_0$ (Hz)
& $0.01^{\dagger}$ & $0.01^{\dagger}$ &  $0.01^{\dagger}$  & $0.01^{\dagger}$\\
&  FWHM$_0$ (Hz)
& $8.5 \pm 1.3$ & $35.53 \pm 10.3$ & $0.063 \pm 0.062$ & $< 9.5 $\\
& Norm$_0$ ($\times10^{-2}$)
& $3.8 \pm 0.5$ & $4.8 \pm 0.9$ & $0.14 \pm 0.042$ & $11\pm 7$\\
\hline

{\texttt{Lorentzian 2}}
& $\nu_1$ (Hz)
& $12.5 \pm 0.7$ & $0.01^{\dagger}$ & $21.7 \pm 2.15$ & $21.4 \pm 2.01$\\
&  FWHM$_1$ (Hz)
& $6.8 \pm 3.0$ & $> 540$ & $38.93 \pm 7.7$ & $40.4 \pm 6.35$\\
& Norm$_1$ ($\times10^{-2}$)
& $1.6 \pm 0.6$ & $26 \pm 18$ & $0.44 \pm 0.049$ & $0.7 \pm 0.09$\\
\hline

{\texttt{Lorentzian 3}}
& $\nu_2$ (Hz)
& $26.5 \pm 2.2$ & --- & $99.7 \pm 2.8$ & --- \\
&  FWHM$_2$ (Hz)
& $11.7 \pm 7.1$ & --- & $16.8 \pm 8.35$ & --- \\
& Norm$_2$ ($\times10^{-2}$)
& $1.2 \pm 0.6$ & --- & $0.13 \pm 0.043$ & --- \\
\hline

{\texttt{Lorentzian 4}}
& $\nu_3$ (Hz)
& $475 \pm 22$ & --- & $422 \pm 130$ &  $130.8 \pm 4.7$\\
&  FWHM$_3$ (Hz)
& $280 \pm 80$ & --- & $< 1000 $ & $< 23$\\
& Norm$_3$ ($\times10^{-2}$)
& $12.6 \pm 2.8$ & ---& $1.2 \pm 0.04$ & $0.06 \pm 0.02$\\
\hline

{\texttt{Lorentzian 5}}
& $\nu_4$ (Hz)
& $146 \pm 11$ & --- & $140.2 \pm 17.3$ & $<439$\\
&  FWHM$_4$ (Hz)
& $124 \pm 41$ & --- & $< 59$ & $1708 \pm 661$\\
& Norm$_4$ ($\times10^{-2}$)
& $6.5 \pm 1.6$ & --- & $0.045 \pm 0.036$ & $4.7 \pm 0.8$\\
\hline


{\texttt{Reduced $\chi^2$} (DOF)}
&  & $0.94$ (82) & $0.94$ (58) & $1.58$ (101) & $0.84$ (109) \\
\hline

\end{tabular}

\label{Table_pds2}
\end{table*}


\subsection{4U 0614+091}

\begin{figure}
        \includegraphics[width=0.5\textwidth]{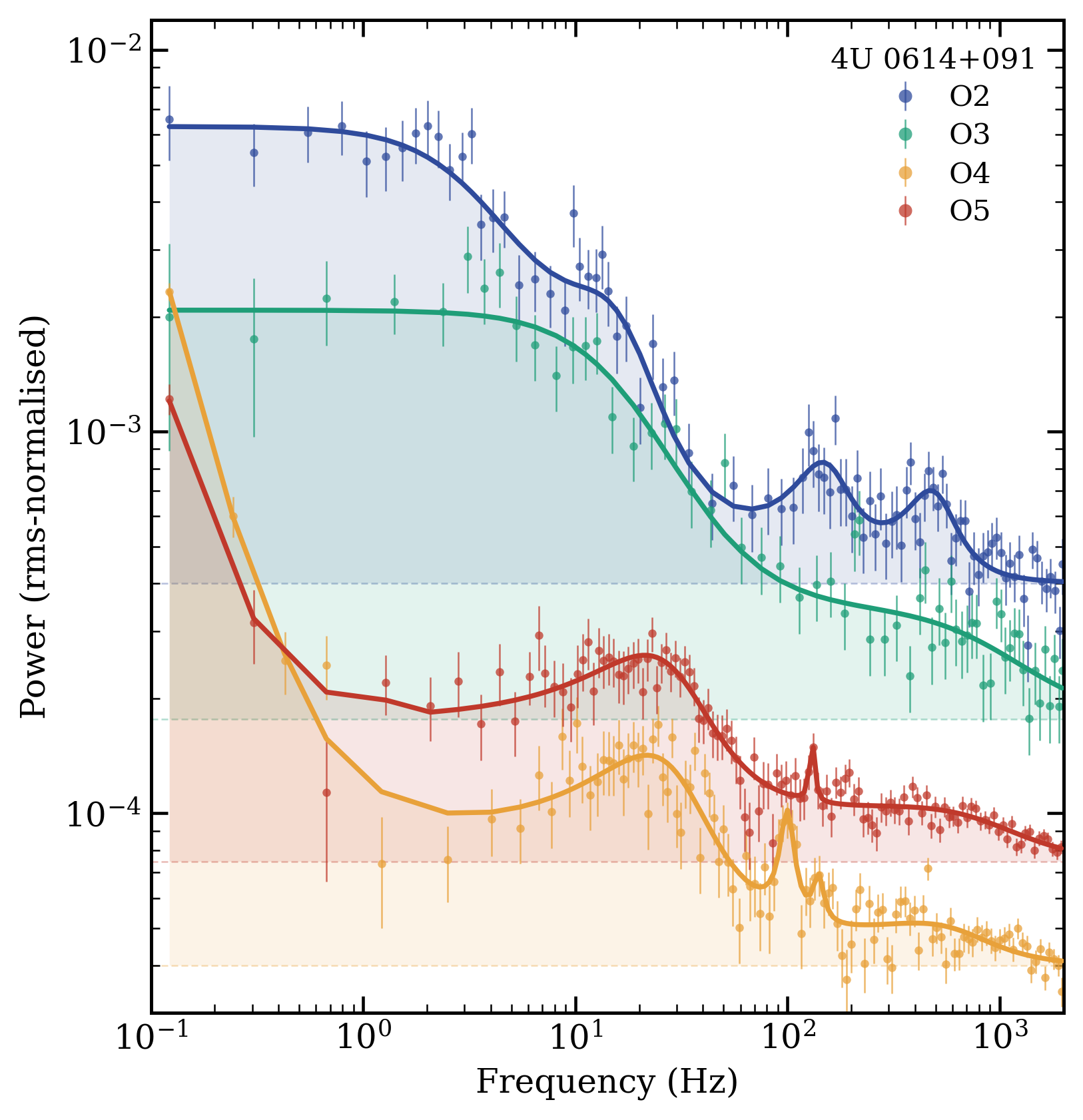}
\caption{Rms-normalised PDS for Observations O2, O3, O4, and O5 of 4U~0614+091,  plotted together on a single set of axes for direct comparison. Data points and best-fitting total models are shown in blue, teal-green, amber, and red for O2, O3, O4, and O5, respectively; each PDS is fitted with multiple Lorentzian components plus a constant. The shaded region beneath each total model indicates the combined contribution of the Lorentzian components above the corresponding fitted constant, which is marked by the dashed horizontal line in the same colour.}
\label{Fig_pds2}
\end{figure}

On the other hand, for 4U 0614+091, a similar extraction procedure has been followed. We have rebinned the raw PDS using the default parameters of the routine "{\textit{laxpc\_rebin\_power powfiles}}". The rebinned PDS were further fitted using the multiple Lorentzian functions. The best-fitted parameters are given below in Table \ref{Table_pds2}. The PDS shown in Figure \ref{Fig_pds2} are rms normalised. Although for observation 3 (see figure \ref{Fig_pds2} green points\color{black}) we do not observe any significant variability in the PDS, we have observed a significant change and appearance of the variabilities in different frequencies in observations 2, 4, 5 (see figure \ref{Fig_pds2} blue, amber and red). In observations 4 and 5, we observe a prominent hectohertz oscillation at $\sim$ 100 Hz, whereas this feature is completely absent in observation 3 but is broad in observation 2. It shows how PDS behaviour changes as the system transitions, making these observations fruitful for studying PDS behaviour alongside spectral changes. kHz QPO is observed in observation 2 as a peak around $\sim 475$ Hz, which is completely absent in the other cases.

\section{Spectral Analysis}
\label{sec:spec}

We have performed joint SXT-LAXPC spectral fitting for each scenario. It is to be noted that for each case, a gain fit with a fixed slope has been added to the SXT data, and a 3\% systematic has been used in all the spectral fittings.

\subsection{1A 1246-588}

\begin{figure}[h]
    \centering
    \includegraphics[width=0.5\textwidth]{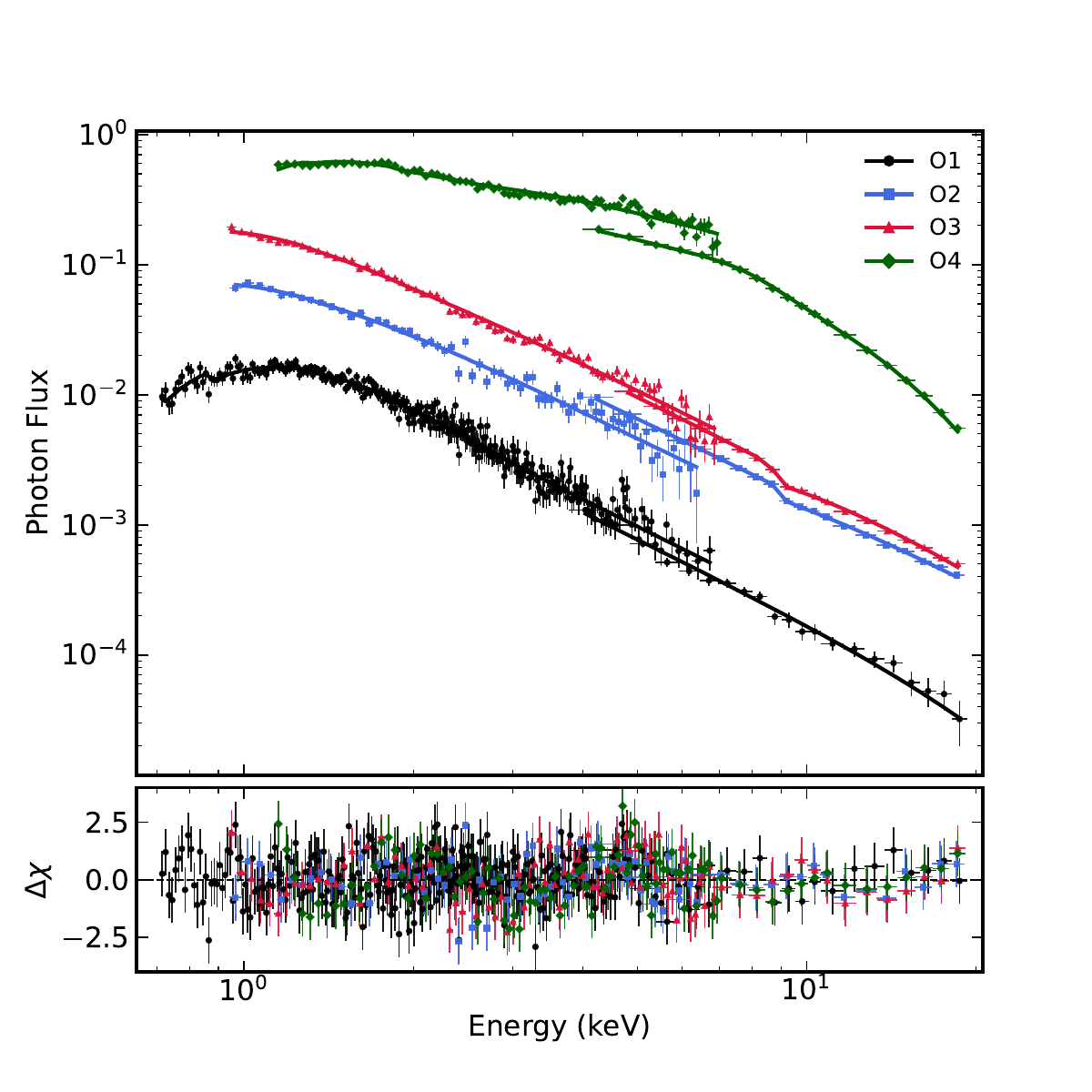}\\
    \caption{Comparison of the spectra of the two different sources A) 1A 1246-588, B) 4U 0614+091. The spectra have been plotted according to the observation number shown in the Table \ref{bpl}. O2-O4 represent the 4U 0614+091 while O1 is representing 1A 1246-588. Photon flux is in units of photons cm$^{-2}$ s$^{-1}$ keV$^{-1}$.\color{black} For details, please see the section \ref{sec:spec}.}
    \label{spectral-plot}
\end{figure}

The spectrum was fitted within 0.7--20 keV for this case. The spectrum is best described by {\texttt{Const * Tbabs * (Thcomp* diskbb)}} (see Figure~\ref{spectral-plot}). ``{\texttt{Tbabs}}'' takes care of the galactic absorption component while ``{\texttt{Thcomp}}'' \citep{2020MNRAS.492.5234Z} represents the thermal Comptonizing component, which takes seed photons from the inner disk component, which is expressed using the multicolour ``{\texttt{diskbb}}'' \citep{1984PASJ...36..741M} component. We did not observe any indication of excess residual emission around 5--7 keV in this scenario. Hence, no additional Gaussian component has been introduced in this scenario. It is further noted that during spectral fitting, we were unable to constrain the system's electron temperature; hence, we set it to its maximum value. The best-fitting parameters have already been noted in Table \ref{tab:spectral_results}. All the error values are reported in the table is calculated at 90\% confidence limit. The disk normalisation is $260^{+170}_{-90}$. The disk normalisation is connected to the inner disk radius through the following relation, $N_{dbb} = \left(\dfrac{R_{in}}{D_{10}}\right)^2 \cos\theta$\footnote{\url{https://heasarc.gsfc.nasa.gov/docs/software/xspec/manual/node165.html}} where $R_{in}$ is the inner disk radius, $D_{10}$ is the distance in units of 10 kpc and $\theta$ is the disk inclination angle. Using a distance of 4.3 kpc \citep{IntZand2008:2008A&A...485..183I}, inclination of $\sim$ 30 degrees ((earlier studies indicate that the source must have very low inclination $<$ 30 degrees, as no significant dips and eclipses have been observed)), and a colour correction factor of 1.7 \citep{1995ApJ...445..780S}, we have calculated the inner disk radius to be around $\sim$20 km, which is plausible for an NSLMXB system given that the neutron star has a theoretical radius of 10 km. Given that the photon index of the system is $\sim$ 2.12, this also supports the disk being close to the compact object, the system being in the soft state, and the spectra being mostly soft-component dominated. This similar behaviour can be observed in many black hole binary systems and NSLMXBs as well.

It should further be noted that the inclusion of an additional soft component, \texttt{bbodyrad}, does not lead to any significant improvement in the fit statistics and appears to be redundant. We therefore explored an alternative model combination, \texttt{const*tbabs*(thcomp*bbodyrad)}, which fundamentally alters the nature of the seed-photon source supplied to the Comptonization component. This model also provides an acceptable fit to the data, yielding a reduced $\chi^2$ of 0.97. The inferred blackbody normalisation is $2304 \pm 1348$, corresponding to a blackbody radius of approximately 21 km. All other spectral parameters remain broadly consistent with those obtained using the previous model configuration. However, since the \texttt{const*tbabs*(thcomp*diskbb)} model yields a lower reduced $\chi^2$ value of 0.91, we adopt the disk-seed-photon configuration for the present analysis. We emphasise that this preference is based solely on the relative goodness of fit and should not be interpreted as evidence favouring a particular physical scenario since we note that the spectral parameters obtained from the two model configurations are mutually consistent within their uncertainties, indicating that the present dataset does not strongly discriminate between a disk-blackbody and a single-temperature blackbody as the seed-photon source for Comptonization.

\subsection{4U 0614+091}

In this scenario, we first explored the simplest model combinations, namely \texttt{const*tbabs*(thcomp*diskbb)} and \texttt{const*tbabs*(thcomp*bbodyrad)}. While both models provide statistically acceptable fits to the observed spectrum, the inferred normalisation values correspond to unrealistically large emitting regions. For the \texttt{bbodyrad} model, the best-fitting normalisation is of the order of $10^{5}$, implying a blackbody radius of approximately 121 km. Similarly, the \texttt{diskbb} model yields a normalisation of $\sim 3.81 \times 10^{4}$, corresponding to a colour-corrected inner disk radius of approximately 255 km according to the prescription described earlier. The inferred emission radii are substantially larger than those typically associated with thermal emission from the neutron-star surface. In addition, the large inner disk radius inferred from the \texttt{diskbb} normalisation may indicate that the simple model does not adequately capture the underlying spectral complexity. Therefore, despite their acceptable statistical quality, these simple model combinations are unlikely to provide a physically self-consistent description of the spectrum.

The other two model combinations that we have further used: (A.) \texttt{Const * edge * Tbabs * (thcomp * diskbb + bbodyrad)} (B.) \texttt{Const * edge * Tbabs * (thcomp * bbodyrad + diskbb)} (see Figure~\ref{spectral-plot}). While models A and B both fit the data well, all values obtained from model A were unacceptable physically. In the case of Model A, the disk normalisation comes out to be of the order of $\sim 2.3 \times 10^4$ (in hard state) while the blackbody normalisation has been kept frozen at 1111, taking a distance of 3.2 kpc, which gives us the radius to be 10 km. Keeping this parameter free makes it unconstrained. While keeping this fixed, we cannot constrain the electron temperature either, as it is pegged at its maximum value of 150 keV. Moreover, the blackbody temperature is also becoming unconstrained.

Best-fitting spectral parameters obtained using Model B are listed in Table~\ref{tab:spectral_results}. In two of the observations corresponding to the soft state, an additional Gaussian component is required to adequately describe the spectrum. The total flux ($F_T$) and disk flux ($F_D$) were calculated using the \texttt{cflux} convolution model in the 0.5--30 keV energy range. All fluxes reported in Table~\ref{tab:spectral_results} are unabsorbed. For 4U~0614+091, a distance of 3.2 kpc and a disk inclination of $62^\circ$ were assumed when deriving the physical disk parameters.

The spectral evolution from O2 to O5 reveals a clear transition from a hard to a soft accretion state. During O2, the spectrum is dominated by Comptonized emission, characterised by a high electron temperature ($\sim$19 keV) and a relatively small disk contribution ($\sim$18\% of the total flux), indicating that inverse Compton scattering is the primary radiative process. As the source evolves to O3, the electron temperature decreases substantially ($\sim$8 keV), while the disk temperature increases and the disk flux fraction rises to $\sim$30\%. This evolution suggests an increasing contribution from the optically thick accretion disk and enhanced cooling of the Comptonizing medium.

In O4, the electron temperature decreases further to $\sim$2.3 keV, while the disk contributes nearly 50\% of the total flux, indicating that the source has entered a pronounced soft state. A notable feature of O4 and O5 is the appearance of an additional Gaussian emission line centred at $\sim$6.97 keV, which is not statistically required in the harder states. The emergence of this feature is consistent with Fe-line emission from the inner accretion flow and may indicate an increased contribution from reflection-related processes during the soft state. The simultaneous presence of a strong disk component, a cool Comptonizing medium, and Fe-line emission suggests a geometry in which the accretion disk plays an increasingly dominant role in shaping the observed spectrum.

Overall, the evolution from O2 to O5 is consistent with progressive cooling of the Comptonizing corona (see figure \ref{kte_vs_Tin}), an increasing contribution from the optically thick accretion disk, and the appearance of spectral features commonly associated with disk reflection in the soft state.

\begin{table*}[ht]
\renewcommand{\arraystretch}{1.5}
\centering
\caption{
Best-fitting spectral parameters obtained using the model \texttt{constant*TBabs*(thcomp*bbodyrad + diskbb)} for four time segments. Parameters marked with $\dagger$ were held fixed during the fitting. Quoted uncertainties correspond to 90\% confidence intervals. Total flux (F$_T$) and disk flux (F$_D$) have been calculated using the {\texttt{cflux}} routine within the 0.5-30 keV energy range. The flux reported in the table is the unabsorbed flux.
}
\begin{tabular}{l c c c c c c c}
\hline
Component & Parameter & Unit & O1 & O2 & O3 & O4 & O5\\
\hline

\texttt{Constant}
& $C_{\rm LAXPC}$ &
& $0.79 \pm 0.03$ & $1.43 \pm 0.11$ & $0.91 \pm 0.04$ & $0.62 \pm 0.02$ & $0.62 \pm 0.02$ \\
\hline

\texttt{TBabs}
& $N_{\rm H}$ & $10^{22}$ cm$^{-2}$
& $0.41^{\dagger}$ & $0.21^{\dagger}$ & $0.21^{\dagger}$ & $0.21^{\dagger}$ & $0.21^{\dagger}$\\
\hline

\texttt{thcomp}
& $\Gamma_{\tau}$ &
& $2.12 \pm 0.08$ & $2.07 \pm 0.03$ & $2.08 \pm 0.06$ & $1.87 \pm 0.08$ & $2.09 \pm 0.06$\\
& $kT_e$ & keV
& $6.0^{\dagger}$ & $19 \pm 15$ & $7.92 \pm 1.91$ & $2.30 \pm 0.09$ & $2.58 \pm 0.09 $\\
& $f_{\rm cov}$ &
& $0.46 \pm 0.06$ & $1.0^{\dagger}$ & $1.0^{\dagger}$ & $1.0^{\dagger}$ & $1.0^{\dagger}$ \\
\hline

\texttt{diskbb}
& $T_{\rm in}$ & keV
& $0.39 \pm 0.04$ & $0.25 \pm 0.05$ & $0.33 \pm 0.04$ & $0.44 \pm 0.03$ & $0.43\pm 0.02$ \\
& $N_{\rm disk} \times10^4$ &
& $0.026^{+0.017}_{-0.009}$ & $0.27^{+0.67}_{-0.17}$ & $0.23^{+0.26}_{-0.14}$ & $0.33^{+0.09}_{-0.13}$ & $0.27^{+0.049}_{-0.061}$ 
\\
\hline

\texttt{bbodyrad}
& $T_{\rm bb}$ & keV
& -- & $0.38 \pm 0.01$ & $0.46 \pm 0.01$ & $0.64 \pm 0.01$ & $0.64 \pm 0.004$\\
& $N_{\rm bb}$ &
& -- & $1111^{\dagger}$ & $1111^{\dagger}$ & $1111^{\dagger}$ & $1111^{\dagger}$ \\
\hline

\texttt{Gaussian}
& $E_{\rm bb}$ & keV
& -- & --- & --- & $6.97^\dagger$ & $6.97^\dagger$\\
& $Width_{\rm g }$ & keV
& -- & --- & --- & $<1.4$ & $0.64 \pm 0.4 $\\
& $N_{\rm g }$ & in $\times 10^{-2}$
& -- & --- & --- & $<0.4 $ & $0.4 \pm 0.2$ \\

\hline

\texttt{flux}
& $F_{\rm T} \times 10^{-9}$ & ergs-cm$^2$-s$^{-1}$
& $0.182 \pm 0.002$ & $0.62 \pm 0.018$ & $1.25 \pm 0.02$ & $3.818 \pm 0.04$ & $3.34 \pm 0.03$\\
& $F_{\rm D} \times 10^{-10}$ & ergs-cm$^2$-s$^{-1}$
& $0.84 \pm 0.01$ & $1.12 \pm 0.09$ & $3.8 \pm 0.15$ & $19.7 \pm 0.31$ & $13.9 \pm 0.24$\\
& Ratio & $\dfrac{F_{\rm D}}{F_{\rm T}}$
& $0.46 \pm 0.01$ & $0.18 \pm 0.02$ & $0.30 \pm 0.01 $ & $0.50^{+0.01}_{-0.01}$ & $0.42 \pm 0.01$\\
\hline

\texttt{Fit quality}
& Reduced $\chi^2$ (dof) &
& $0.91$ (316) & $0.75$ (95) & $1.04$ (93) & $1.14$ (103) & $0.72$ (100) \\
\hline

\end{tabular}

\label{tab:spectral_results}
\end{table*}

\section{Summary \& Discussions}

In this work, we present a broadband spectro-temporal study of the UCXB candidates 1A~1246$-$588 and 4U~0614$+$091 using {\em AstroSat}, and compare the results with our earlier findings reported on NSLMXBs. The primary aim is to investigate the connection between spectral evolution and rapid X-ray variability in these systems. For 1A~1246$-$588, we detect a narrow upper kHz QPO at $993 \pm 4$ Hz with a width of $<12$ Hz and a significance of $3.7\sigma$, observed during a relatively soft spectral phase. An independent cross-instrument (co-spectrum) check did not confirm this feature (Section~\ref{sec:temp}), and given its already marginal significance we treat it as a tentative, unconfirmed signature rather than a secure detection. In contrast, no significant kHz QPO is detected from 4U~0614$+$091. Instead, the source shows low-frequency and hecto-Hz QPOs as it evolves toward softer spectral states. In contrast, no significant kHz QPO is detected from 4U~0614$+$091. Instead, the source shows low-frequency and hecto-Hz QPOs as it evolves toward softer spectral states.

\subsection{Similarities between the two systems}

Despite these differences in timing behaviour, both systems show spectra that can be described by thermal Comptonization coupled to a soft thermal component, with physically plausible emitting radii indicating that the dominant thermal emission originates close to the neutron star. The soft-state spectra in both sources are characterised by a stronger thermal contribution and a reduced role for Comptonization, consistent with enhanced cooling of the hot electron cloud due to an increasing supply of soft seed photons.

\subsection{What is different in 4U 0614+091?}

The most important distinction between the two systems is the extent of their spectral evolution. While 1A~1246$-$588 appears to remain predominantly in a soft accretion state throughout the observation, 4U~0614$+$091 undergoes a clear hard-to-soft transition, allowing the evolution of the accretion flow to be tracked directly. In 4U~0614$+$091, the electron temperature decreases from $\sim19$ keV to $\sim2.3$ keV, while the disk contribution increases from $\sim18\%$ to $\sim50\%$ of the total flux. Furthermore, a Gaussian emission feature near 6.97 keV emerges in the soft state, suggesting the onset of reflection-related processes that are not observed in 1A~1246$-$588. We also find that simple single-component thermal models yield physically implausible emitting radii for 4U~0614$+$091, necessitating a more complex spectral geometry, whereas the corresponding models for 1A~1246$-$588 provide physically reasonable parameters. These results indicate that, while both systems share the general spectral characteristics of soft-state UCXBs, 4U~0614+091, for which multiple observations spanning distinct hard and soft states are available, exhibits substantially richer spectral evolution and provides direct evidence for changes in accretion geometry across state transitions. Since the single available observation of 1A~1246$-$588 corresponds to only one (soft) spectral state, no state-transition-related evolution can be assessed for this source with the present dataset, and we therefore do not draw any conclusion regarding its accretion geometry during state transitions.

\subsection{Similarities and differences with the previous studies}

\begin{figure}[t]
\centering
\includegraphics[width=0.95\columnwidth]{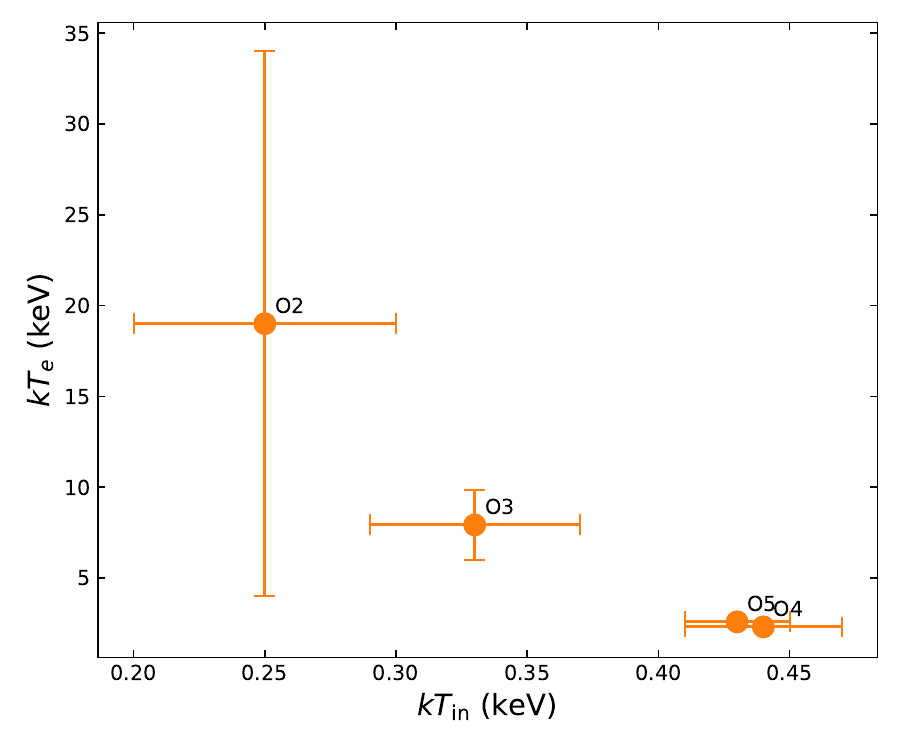}

\vspace{0.2cm}

\includegraphics[width=0.95\columnwidth]{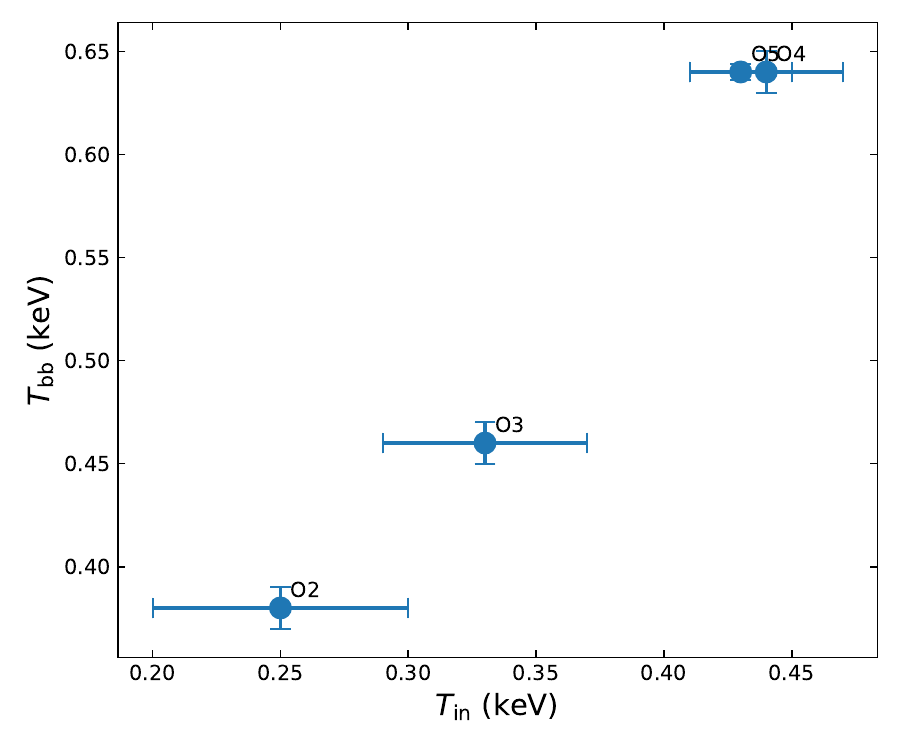}

\caption{Relation between the spectral parameters for O2-O5 in 4U~0614+091:
Top: Electron temperature as a function of inner disk temperature.
Bottom: Blackbody temperature as a function of inner disk temperature.
}
\label{kte_vs_Tin}
\end{figure}

Recently, \citet{moutard2023simultaneousnicernustarobservations} investigated the accretion geometry of 4U~0614+091 using simultaneous {\em NICER} and {\em NuSTAR} observations and relativistic reflection modelling. They found that the accretion disk generally extends close to the neutron star ($\sim6\,R_g$), becoming mildly truncated at $\sim11.5\,R_g$ during the lowest-flux observation. In our analysis, the inner disk radius inferred from the \texttt{diskbb} normalisation remains within a relatively narrow range of $\sim13$--$15\,R_g$ across O2--O5. Although the radii derived from continuum modelling are systematically larger than those obtained from relativistic reflection spectroscopy (Tables 4 and 5 in \citet{moutard2023simultaneousnicernustarobservations}), they are broadly consistent with a moderately truncated accretion disk.

The spectral evolution appears to be driven primarily by changes in the thermal properties of the accretion flow rather than by substantial variations in the inner disk radius. In particular, the inner disk temperature increases from $\sim0.25$ keV in O2 to $\sim0.44$ keV in O4, while the electron temperature decreases from $\sim19$ keV to $\sim2.3$ keV. This behaviour is accompanied by an increase in both the disk flux and the disk-flux fraction, indicating enhanced cooling of the Comptonizing corona by soft seed photons originating from the disk. Furthermore, the blackbody temperature associated with the neutron-star surface and/or boundary layer increases from $\sim0.38$ keV to $\sim0.64$ keV and exhibits a positive correlation with the disk temperature (see row 2 of Figure \ref{kte_vs_Tin}). This coupled increase in $T_{\rm bb}$ and $T_{\rm in}$ suggests that both the accretion disk and the boundary layer respond to an increasing mass accretion rate, leading to a hotter thermal component while simultaneously cooling the corona. Therefore, while the accretion disk remains approximately at a similar radius throughout the observations, substantial changes occur in the energetics and thermal coupling of the disk--corona--boundary-layer system.

A recent study by \citet{Poojyam2026} presented a detailed spectral investigation of 1A~1246$-$588 using {\em NICER} and {\em AstroSat} observations and showed that the source follows an atoll-like evolution, with the spectral changes primarily driven by a redistribution of accretion power between the thermal boundary layer and the Comptonizing region. Their broadband spectral modelling favoured a blackbody plus Comptonization continuum and did not require a statistically significant multicolour disk component, leading them to interpret the soft emission as originating predominantly from the neutron-star surface or boundary layer. In contrast, our {\em AstroSat} analysis requires a disk component in addition to thermal Comptonization, suggesting that the accretion disk extends close to the neutron star. Furthermore, the difference between the two studies is not limited to the choice of the soft component. \citet{Poojyam2026} employed the \texttt{nthcomp} model, whereas we use the \texttt{thcomp} model. Since these Comptonization models differ in their treatment of the seed-photon distribution and the resulting Comptonized photon spectrum, the inferred spectral decomposition and the relative contributions of the thermal and Comptonized components may differ accordingly. Nevertheless, both studies consistently highlight the central role of the Comptonizing region in shaping the source's spectral evolution.

\subsection{Connection to the temporal variabilities}

The PDS of O2 exhibits the richest variability among all the observations, with significant components detected from $\sim12$ Hz up to $\sim1$ kHz. Two low-frequency Lorentzians are observed at $12.5\pm0.7$ Hz and $26.5\pm2.2$ Hz. The frequency ratio of these components is close to two, suggesting a possible fundamental--harmonic relationship. At higher frequencies, broad features are detected at $146\pm11$ Hz, and $475\pm22$ Hz. The presence of the $\sim475$ Hz component is noteworthy, as its frequency falls within the range typically associated with kHz QPOs in NSLMXBs. While the broad nature of these features prevents a secure classification, they indicate the presence of kHz-QPO activity at similar frequencies previously reported by {\em NICER} \citep{2018ApJ...860L...9B}.

The timing properties of O4 are characterised by three distinct variability components. A broad low-frequency oscillation is detected at $21.7\pm2.2$ Hz, consistent with the $\sim21$ Hz feature also observed in O5 and possibly related to the low-frequency variability seen in O2. In addition, O4 exhibits a relatively narrow and significant feature at $99.7\pm2.8$ Hz, which is unique to this observation and represents the most prominent peaked component in the power spectrum. A broader component is also detected at $140.2\pm17.3$ Hz. Interestingly, variability near $\sim140$ Hz is present in O2, O4, and O5, despite the substantial differences in their spectral properties, suggesting the presence of a characteristic timescale that remains relatively stable across different source states.

Such behaviour suggests that while the thermal evolution of the disk--corona system strongly influences the overall variability properties, some characteristic variability timescales may remain relatively insensitive to changes in the spectral state. The detection of broad high-frequency feature at $\sim475$ Hz in O2 further indicates enhanced rapid variability when the Comptonizing region is energetically dominant in 4U 0614+091, consistent with the general picture observed in atoll sources where stronger hard emission is often associated with richer high-frequency timing behaviour \citep{soma-2025,vanderKlis2006}. Although this can not be generalized for the 1A 1246-588 depending only on the single observation that we have in our hand for this present situation.

In contrast, O4 and O5 are characterised by hotter thermal components, a substantially cooler corona, and comparatively simpler timing behaviour. Such a coupling between spectral state and rapid variability has been reported in several NSLMXBs, where the timing properties evolve systematically with changes in the accretion geometry and coronal conditions \citep{vanderKlis2006,vanderKlis2010}. The increase in disk and boundary-layer temperatures accompanied by a decrease in the coronal temperature suggests enhanced cooling of the Comptonizing region by soft seed photons, while the reduction of high-frequency variability in the softer observations may indicate a diminished role of the corona in generating rapid X-ray fluctuations.

The broad feature detected at $\sim475$ Hz in O2 further suggests the presence of kHz-QPO activity. High-frequency oscillations in atoll sources are generally believed to originate in the innermost regions of the accretion flow, close to the neutron star and the inner disk edge \citep{vanderKlis2006,Stella1999}. Moreover, spectral studies of neutron-star systems have shown that the appearance and properties of kHz QPOs are closely linked to changes in the Comptonizing corona and the thermal components of the spectrum \citep{Zhang2017,soma-2025}. In particular, \citet{Zhang2017} found that the presence of kHz QPOs is associated with specific coronal conditions during spectral-state transitions, while \citet{soma-2025} reported a strong connection between the evolution of kHz QPOs and variations in the disk, boundary-layer, and coronal parameters. Therefore, the coexistence of strong coronal emission and broad high-frequency variability in O2 is qualitatively consistent with a scenario in which rapid variability is generated within the inner accretion flow and is modulated by the physical state of the corona.

\section{Acknowledgments}

We have utilized the data from LAXPC and SXT payloads onboard {\it AstroSat} available at ISSDC. We wish to acknowledge the LAXPC Payload Operation Center (POC) and SXT POC at TIFR, Mumbai. The research leading to these results has been funded by the Department of Space, Govt. of India, ISRO under grant no. DS\_2B-13013(2)/10/2020-Sec.2. We thank the referee for constructive comments that helped improve this manuscript.


\section{Data Availability}
The LAXPC and SXT archival data that has been used in this article can be found at {\it AstroSat} ISSDC website (\url{https://astrobrowse.issdc.gov.in/astro_archive/archive}).

\bibliographystyle{elsarticle-harv}
\bibliography{paper_5}
\label{lastpage}
\end{document}